# Coaxial Gyromagnetic Nonlinear Transmission Line with a Slow-Wave Structure: 2D/3D FDTD Modeling

Serhii Karelin and Ivan Onishchenko

***Abstract*— A coaxial gyromagnetic nonlinear transmission line (GNLTL) enables pulse-front compression and oscillation packet generation in the 0.3–10 GHz range. Numerical modeling of such systems typically employs a 2D axially symmetric FDTD scheme coupled with the parallel solution of full non-linearized Landau-Lifshitz-Gilbert equations. Previous studies have shown that using an insulating dielectric with a permittivity ε>60 enables the generation of long-lasting oscillations via a Cherenkov-like effect. However, practical implementation is limited due to the severe high-frequency losses inherent in materials with such extreme permittivity, like water. To overcome this limitation and enhance performance, we propose a novel design utilizing a spiral central conductor to mimic a high-permittivity environment, with the ferrite placed inside the spiral. Due to the loss of axial symmetry, the FDTD model was extended to 3D coordinates. Simulations confirm that the spiral conductor configuration generates a prolonged oscillation packet similar to that of a high-permittivity dielectric waveguide, but with significantly reduced losses and improved energy transfer efficiency.**

## I. INTRODUCTION

GYROMAGNETIC nonlinear transmission lines (GNLTLs) have attracted significant interest over the past few decades as a reliable and compact technology for high-voltage pulse-front compression and high-power microwave generation [1–4]. Although the peak of research activity in this field occurred approximately 5–15 years ago, gyromagnetic lines continue to be a subject of intense interest today [5–11]. The resulting high-frequency oscillating waveforms generated by these systems are highly suitable for practical applications, most notably as driving sources for directional antenna radiation and phased arrays [3].

Given that high-voltage experiments in this field are technically complex, costly, and involve significant safety risks, numerical simulation serves as the primary tool for investigating GNLTL dynamics. Furthermore, because these systems are inherently nonlinear, conventional frequency-domain analysis methods offer extremely limited insight, making time-domain numerical modeling the only viable approach for capturing the full complexity of shock-wave formation and oscillation generation. Early numerical studies, pioneered by Dolan [2], utilized a one-dimensional (1D) approximation based on telegrapher's equations coupled with the Landau-Lifshitz-Gilbert (LLG) equation to describe the ferrite's response. To provide a more comprehensive understanding of the electromagnetic field distribution, this approach was subsequently extended to a two-dimensional (2D) axially symmetric model based on the full set of Maxwell's equations [12].

Using this 2D approach, it was subsequently demonstrated that incorporating an insulating dielectric with an extremely high relative permittivity (ε> 60) enables the generation of long-lasting, slowly decaying oscillation trains via a Cherenkov-like radiation mechanism [13]. However, the practical potential of such configurations remains severely constrained due to the high-frequency dielectric losses inherently present in real materials with extreme permittivity, such as water.

To overcome this fundamental physical limitation and enhance system performance without relying on lossy high ε media, this paper proposes a novel design utilizing a spiral central conductor structure to geometrically mimic a high-permittivity environment, with the ferrite core positioned inside the helix. Because the introduction of this spiral configuration inherently breaks the axial symmetry of the line, the established 2D FDTD framework [12] becomes insufficient, necessitating an extension to a fully three-dimensional (3D) Cartesian coordinate system. In this work, we present a complete 3D FDTD analysis of this spiral GNLTL concept, the preliminary idea of which was introduced in [14]. Simulations confirm that the proposed configuration successfully generates a prolonged high-frequency oscillation packet comparable to that of a theoretical high-permittivity dielectric-filled line, while significantly reducing high-frequency dielectric losses.

## II. BASELINE CONCEPT AND 2D AXISYMMETRIC BACKGROUND

### A. 2D Axisymmetric FDTD-LLG Numerical Methodology

The specialized numerical analysis of wave propagation in GNLTLs was historically initiated using 1D analytical and circuit-based models [2]. To capture the real spatial structure of the electromagnetic fields and the non-uniform radial distribution of magnetization, a comprehensive 2D axially symmetric framework was developed [12]. This approach solves the full set of time-dependent Maxwell's equations for

This work was supported in part by the National Academy of Sciences of Ukraine.

S. Karelin and I. Onishchenko are with the National Science Center «Kharkiv Institute of Physics and Technology», Kharkiv 61108, Ukraine (e-mail: sergeykarelin1976@gmail.com; onish@kipt.kharkov.ua)

the field components combined with the full non-linearized LLG equation governing the magnetization vector dynamics within the ferrite core.

At each time step *n*, the computational cycle executes three self-consistent stages:

1. *Electromagnetic Field Update:* Given $H^n$ and $E^n$ at time step *n*, the standard FDTD scheme is used to compute $B^{t+1}$ ($B=H+M$) and $E^{n+1}$ at the next time step.

2. *Nonlinear Magnetization Integration:* Having the magnetization $M^n$ at time n, the Runge–Kutta method is used to compute $M^{n+1}$ at the next time step according to the following scheme:

$\overrightarrow{K1} = h_t L(\vec{H}^{n+1}, \vec{M}^{n+1})$,

$\overrightarrow{K2} = h_t L(\overrightarrow{H2}, \overrightarrow{M2})$,

where $\overrightarrow{M2} = \vec{M}^n + \overrightarrow{K1}, \overrightarrow{H2} = (\vec{H}^n + \vec{B}^{n+1} - \overrightarrow{M2})/2$,

$\overrightarrow{K3} = h_t L(\overrightarrow{H3}, \overrightarrow{M3})$,

where $\overrightarrow{M3} = \vec{M}^n + \overrightarrow{K2}, \overrightarrow{H3} = (\vec{H}^n + \vec{B}^{n+1} - \overrightarrow{M3})/2$,

$\overrightarrow{K4} = h_t L(\overrightarrow{H4}, \overrightarrow{M4})$,

where $\overrightarrow{M4} = \vec{M}^n + \overrightarrow{K3}, \overrightarrow{H4} = \vec{B}^{n+1} - \overrightarrow{M4}$,

$\vec{M}^{n+1} = \vec{M}^n + (\overrightarrow{K1} + 2\overrightarrow{K2} + 2\overrightarrow{K3} + \overrightarrow{K4})/6$,

where $L(\vec{H}, \vec{M}) = \gamma\mu_0[\vec{M} \times \vec{H}] - \alpha \frac{\gamma\mu_0}{M_S}[\vec{M} \times [\vec{M} \times \vec{H}]]$ – right part of LLG equations.

$h_t$ – time step.

3. *Magnetic Field Correction:* $H^{n+1}$ required for the next time step is determined from the updated $B^{n+1}$ and $M^{n+1}$:

$\vec{H}^{n+1} = \vec{B}^{n+1} - \vec{M}^{n+1}$.

This computational framework has demonstrated high efficiency in modeling GNLTL [15]. A similar approach was used by other authors [16].

### *B. High-Permittivity Cherenkov-like Effect*

Previous 2D numerical simulations [13] revealed a unique physical transition when modifying the properties of the insulating dielectric layer. Typically, such lines utilize conventional transformer oil with a low relative permittivity (ε=2.25), which supports stable shock-front sharpening and oscillation generation. However, it was found that as the permittivity of this insulating medium is systematically increased, the standard high-frequency gyromagnetic oscillations initially weaken and eventually disappear, but then reappear as a modified oscillation regime when the relative permittivity reaches an extreme threshold of ε≈60. These newly emerged oscillations exhibit a lower initial peak amplitude but feature an extraordinarily slow decay rate, resulting in a significantly prolonged microwave packet length, as illustrated in Fig. 1.

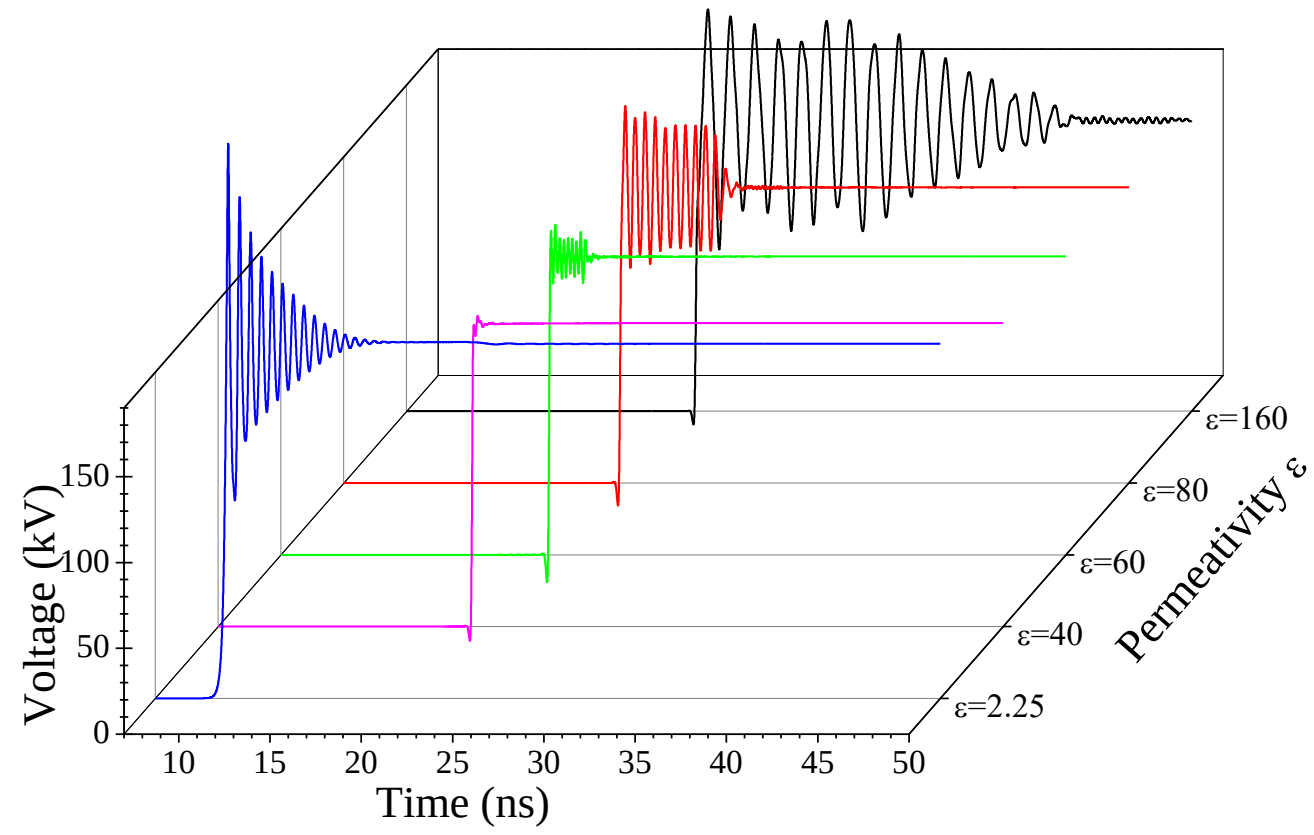

**Fig. 1.** Output voltage waveforms computed via the 2D axisymmetric model for different permittivities of the insulating layer.

The physical mechanism of this phenomenon is evident from the transverse magnetic field distribution in Fig. 2: the shock wave in the ferrite generates a conical wave in the dielectric, analogous to the Cherenkov effect with an angle satisfying sinα=$v_d/v_f$, where $v_d$ is the electromagnetic wave speed in the dielectric and $v_f$ is the speed in the ferrite (see Fig. 2).

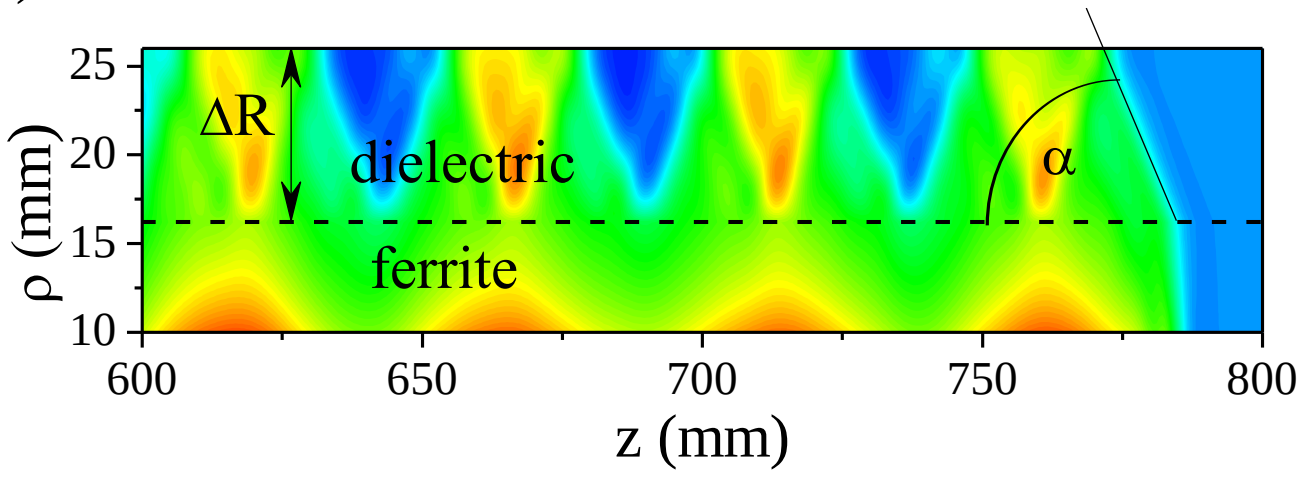

**Fig. 2.** Calculated spatial distribution of the transverse magnetic field within the 2D GNLTL cross-section.

The conical wave subsequently reflects from the waveguide walls, thereby exciting oscillations. Their period is readily determined by a simple formula:

$$T = \frac{4\Delta R\sqrt{v_f^2 - v_d^2}}{v_d v_f} \quad (1)$$

In terms of its observable features and underlying physics, this phenomenon is analogous to the synchronous wave generation observed in discrete nonlinear transmission lines [17–19].

## III. Design Evolution and 3D Model Justification

Thus, the phenomenon mentioned above yields longer and narrower-spectrum oscillations compared to the classical design (ε=2.25), making them more suitable for practical applications, e.g., antenna radiation. However, in practice, only polar liquids exhibit such high permittivity, but their application is severely constrained by high-frequency dielectric losses. Among all polar fluids, conventional purified water represents the only viable candidate due to its low cost, safety, and high static permittivity combined with a relatively high relaxation frequency.

Nevertheless, even for water, this finite relaxation time, as captured by the Debye model, introduces losses that prevent stable RF generation [6]. These losses can be partially mitigated by shifting the oscillation spectrum to lower frequencies, which, according to (1), is achieved by systematically increasing the transverse dimensions of the transmission line structure.

To evaluate this, comparative calculations were performed for the baseline design and its scaled versions by ×2, ×5, ×7, where the transverse radius $R$ and line length $L$ were scaled proportionally. To maintain the same initial magnetic field intensity across the different geometries, the input pulse voltage $U$ was scaled proportionally as well. The water layer was modeled using the standard Debye dispersion framework [10] with parameters: $\varepsilon_0$=78.4, $\varepsilon_\infty$=3.1, and $t_0$=8.27·10$^{-12}$ s. As shown in Fig. 3, increasing the structural dimensions by 5 and 7 times renders dielectric attenuation practically insignificant for oscillation generation. However, fabricating such an oversized GNLTL is technologically unfeasible due to excessive weight, volume, and high-voltage insulation constraints.

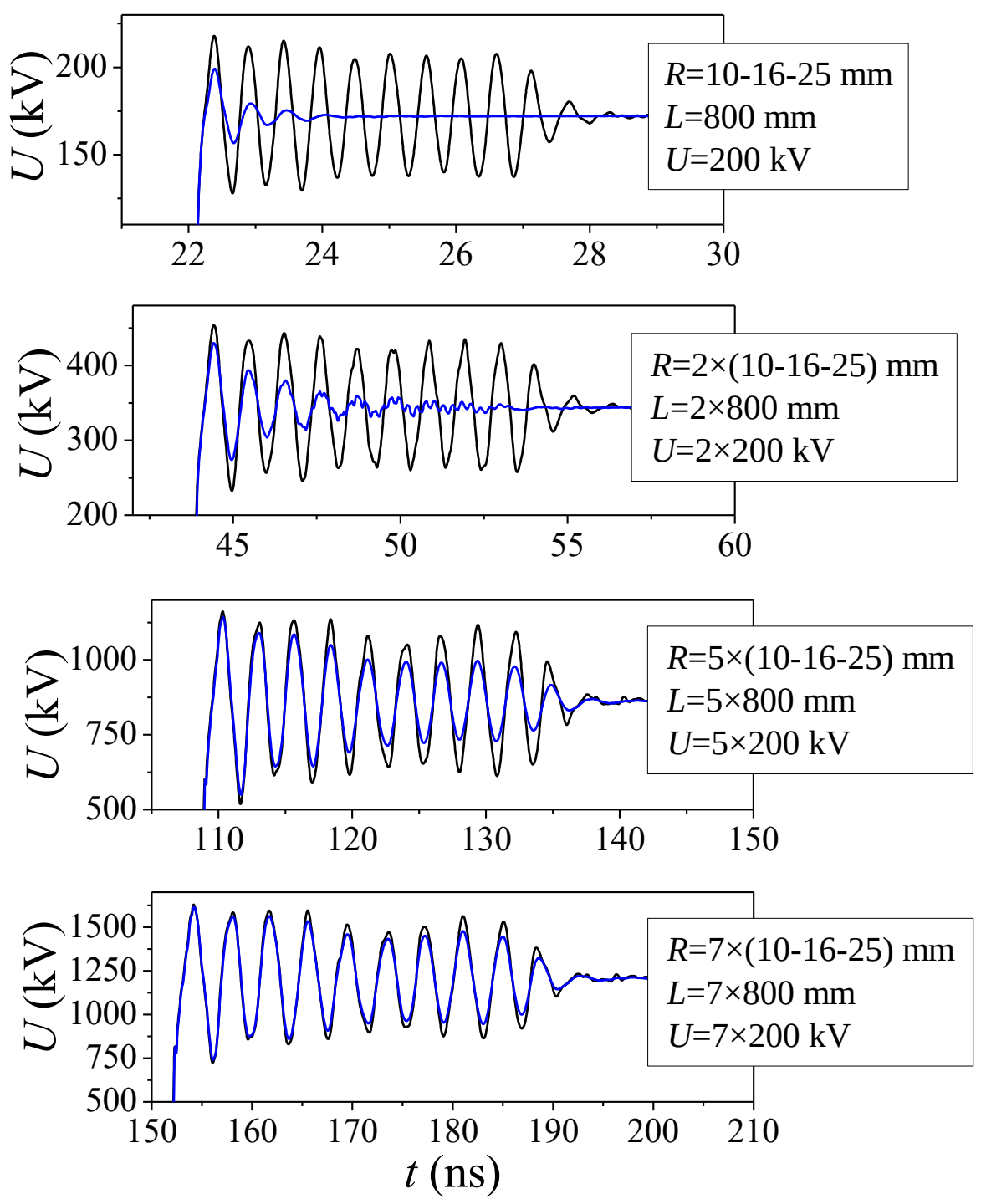

**Fig. 3.** Simulated output voltage waveforms for the baseline and scaled (×2, ×5, ×7) GNLTL designs: black curves correspond to the idealized lossless medium, and blue curves include water losses modeled via the Debye relaxation framework.

Another approach to the practical implementation of this regime is the utilization of a periodic metal disk structure. Numerical simulations and experiments have demonstrated [18–20] that while this method is viable, it remains insufficient for practical applications. Specifically, it either yields a long oscillation train with a low amplitude (~15% of the input pulse voltage) or a higher amplitude (~40%) but with a short duration that offers no improvement over the classical design. The reason for this limitation is straightforward: such a structure decelerates the wave only within a narrow high-frequency range. Therefore, a slow-wave structure capable of decelerating the wave across the entire frequency spectrum, including low frequencies, is required to fully emulate a high-permittivity medium. Essentially, a spiral structure is the only configuration that satisfies this requirement. Therefore, an alternative wave-slowing approach was proposed: a spiral central conductor emulating a high-permittivity material, with the ferrite placed inside the spiral.

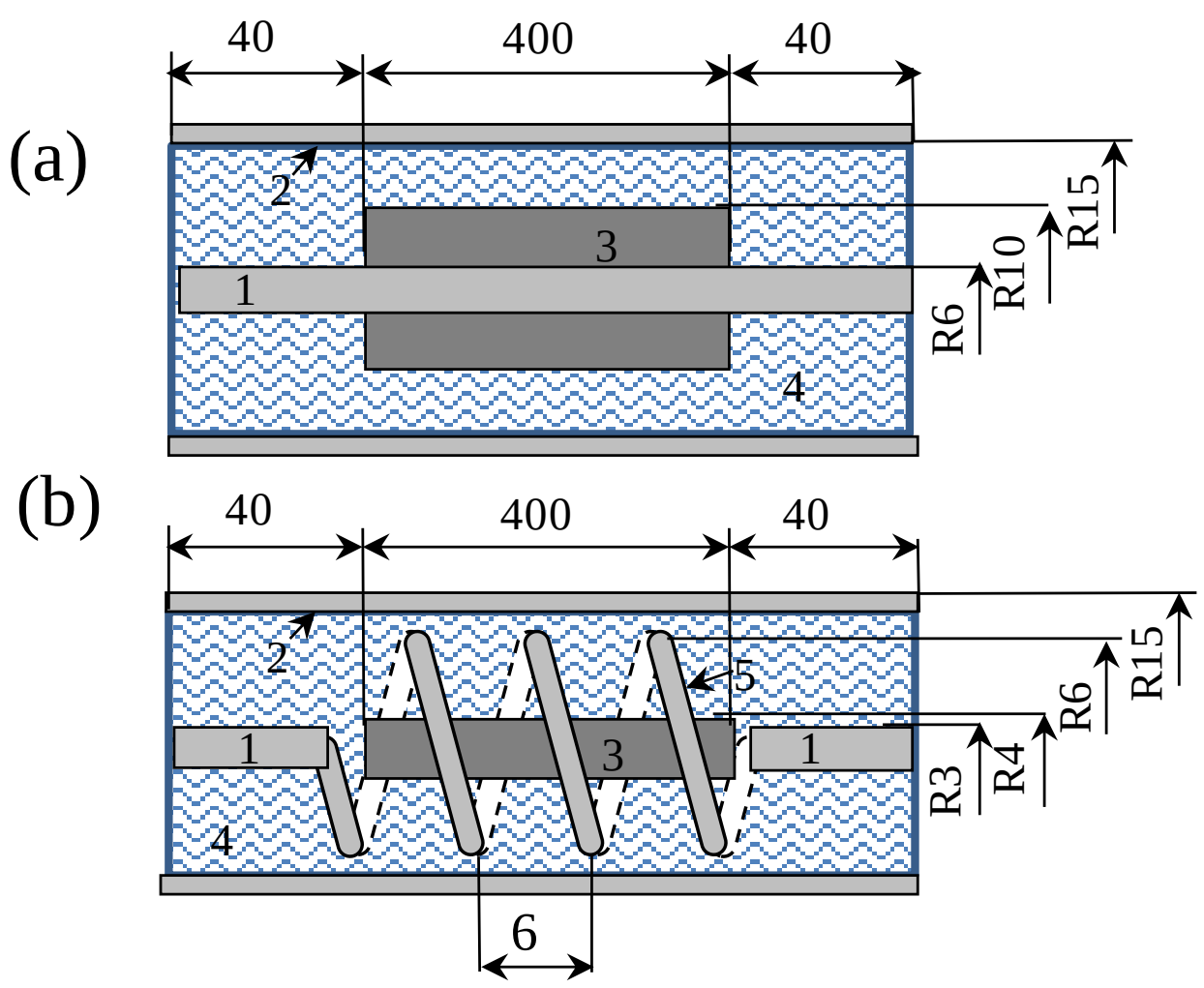

**Fig. 4.** Schematic diagrams of the GNLTL layouts: (a) baseline coaxial design with a solid inner conductor, and (b) proposed alternative configuration with a spiral central conductor. 1 – inner conductor, 2 – outer conductor, 3 – ferrite, 4 – insulating dielectric, 5 – spiral conductor.

## IV. Extension to 3D Modeling and Verification

Due to the loss of axial symmetry in the proposed spiral conductor design, it cannot be modeled by the existing 2D axisymmetric methodology [12], so this approach was extended to a fully 3D numerical framework formulated in a standard Cartesian coordinate system. To establish the numerical validity of the developed 3D framework, a mesh convergence analysis was first conducted using the baseline coaxial design (see Fig. 4(a)) under different spatial grid steps: Δh = 2 mm, 1~mm, and 0.5~mm. As can be seen from Fig. 5(a), the simulation loses numerical stability at Δh = 2 mm, which manifests as atypical, unphysical distortions. For the finer meshes Δh = 1 mm and 0.5 mm, the calculated waveform profiles are qualitatively identical. However, a noticeable 4.5% difference is observed in the signal delay time, defined as the time interval between the peaks of the input and output pulses. Conversely, the period of the first oscillation (the distance between the first and second peaks) remains practically identical in both cases, measuring 0.34 ns.

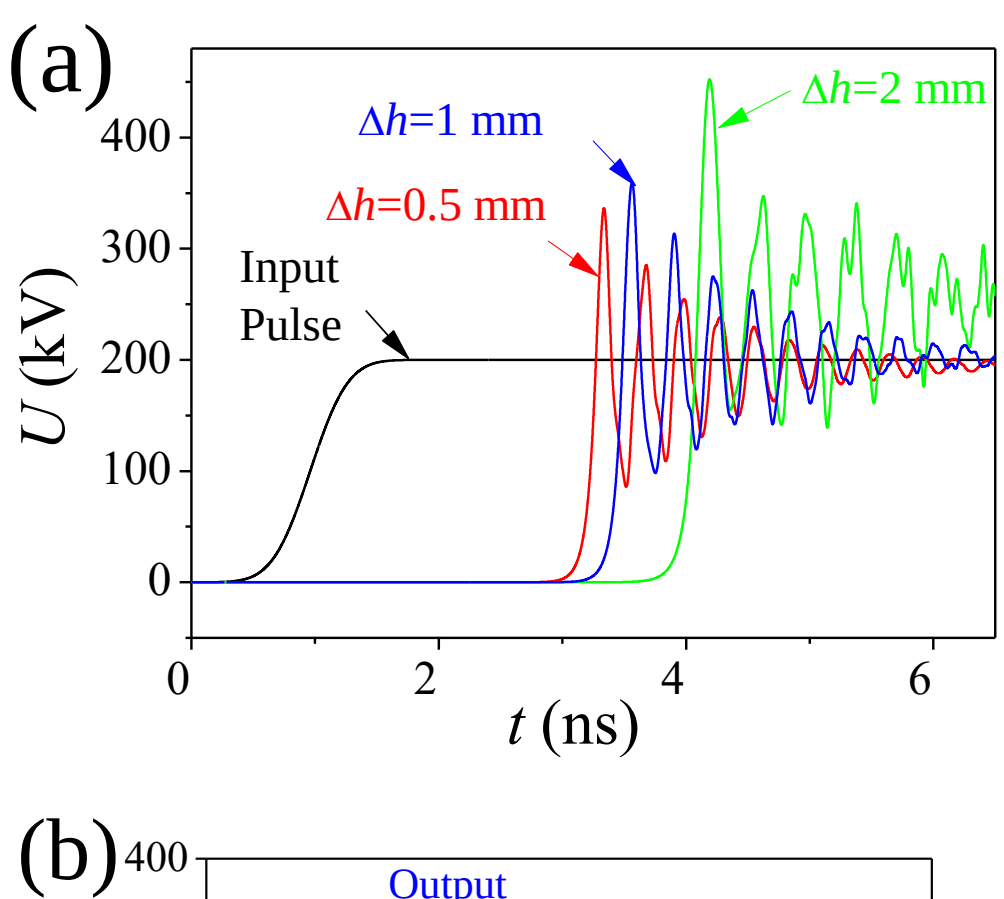

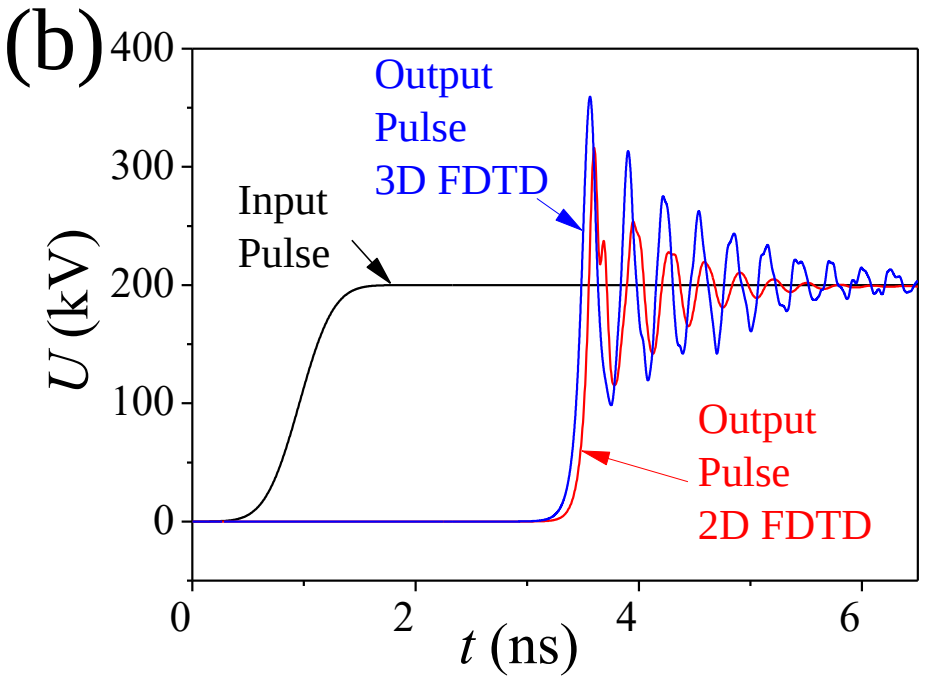

**Fig. 5.** Convergence and verification analysis of the developed 3D computational framework: (a) simulated 3D waveforms under different spatial grid steps (Δh= 2 mm, 1 mm, and 0.5 mm) for the baseline design, and (b) comparison of output voltage waveforms obtained via 2D axisymmetric and 3D modeling using Δh = 1 mm. Ferrite: $M_S$= 300 kA/m, $H_0$=30 kA/m, α=0.1, $\varepsilon_f$=12. Insulating dielectric: $\varepsilon_d$=2.25.

Consequently, the spatial step of Δh=1 mm was selected as an optimal trade-off between computational accuracy and resource efficiency. Using this grid configuration, a comparative simulation was performed against the proven 2D axisymmetric model. As shown in Fig. 5(b), the comparison demonstrates satisfactory agreement: the generated oscillations in the 3D model exhibit nearly the same frequency as in the 2D case, though they feature a marginally higher amplitude.

Thus, although the developed modeling tool possesses an estimated accuracy of approximately 5–10%, it is entirely sufficient for at least a qualitative verification of the operability of the proposed spiral conductor design.

## V. Modeling of the Spiral Design and Discussion

To evaluate the performance and physical advantages of the proposed design, a comprehensive 3D numerical simulation was performed. The computational setup compared three distinct GNLTL configurations under identical input pulse and biasing conditions, with the resulting output voltage waveforms presented in Fig. 6.

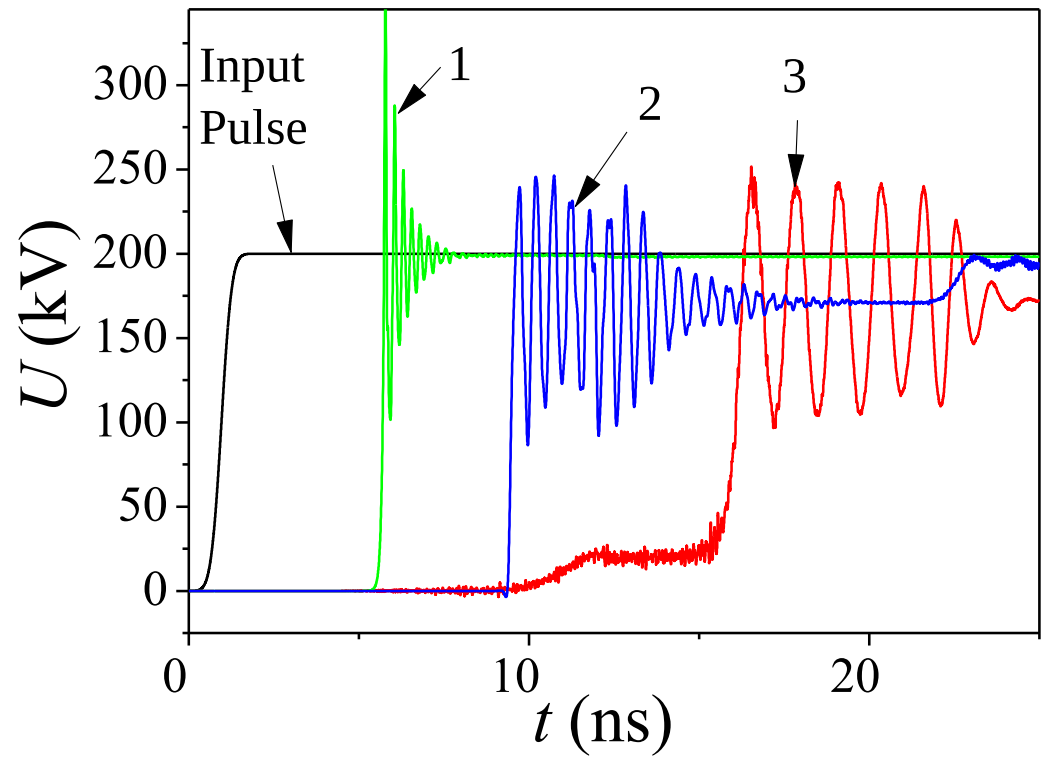

**Fig. 6.** Comparison of simulated output voltage waveforms for different GNLTL configurations: 1 – the classical baseline design with low permittivity ($\varepsilon_d$=2.25), 2– the idealized baseline design with high permittivity ($\varepsilon_d$=80), 3 – the proposed alternative configuration featuring a spiral central conductor ($\varepsilon_d$=2.25).

As can be seen from Fig. 6, the classical GNLTL design filled with a low-permittivity dielectric ($\varepsilon_d$=2.25) produces a standard, rapidly decaying shock-front oscillation. Conversely, the configuration with a high-permittivity medium ($\varepsilon_d$=80) yields a significantly prolonged oscillation train, and the proposed spiral conductor configuration ($\varepsilon_d$=2.25) demonstrates a closely similar behavior, successfully replicating the extended microwave packet length without requiring an actual high-permittivity insulation layer. Furthermore, the simulation indicates that the oscillation amplitude generated by the spiral structure is at least equal to, or marginally higher than, that of the theoretical high-permittivity benchmark.

The transverse profiles of the electric and magnetic fields in the spiral conductor design (Fig. 7) reveal key practical aspects. The electric field magnitude reaches peak values up to ~1000 kV/cm — a level posing significant risk of electrical breakdown. However, the maximum is located on the outer side of the spiral, meaning potential breakdown would not destroy the ferrite core. Conversely, due to the wave slowing effect in the spiral conductor, the magnetic field is strongly concentrated inside the spiral, precisely where the ferrite is located. This concentration is stronger for higher wave slowing. According to the LLG equation, this enhances the ferrite nonlinearity and improves energy coupling between the shock wave and excited wave, thereby increasing the overall generation efficiency.

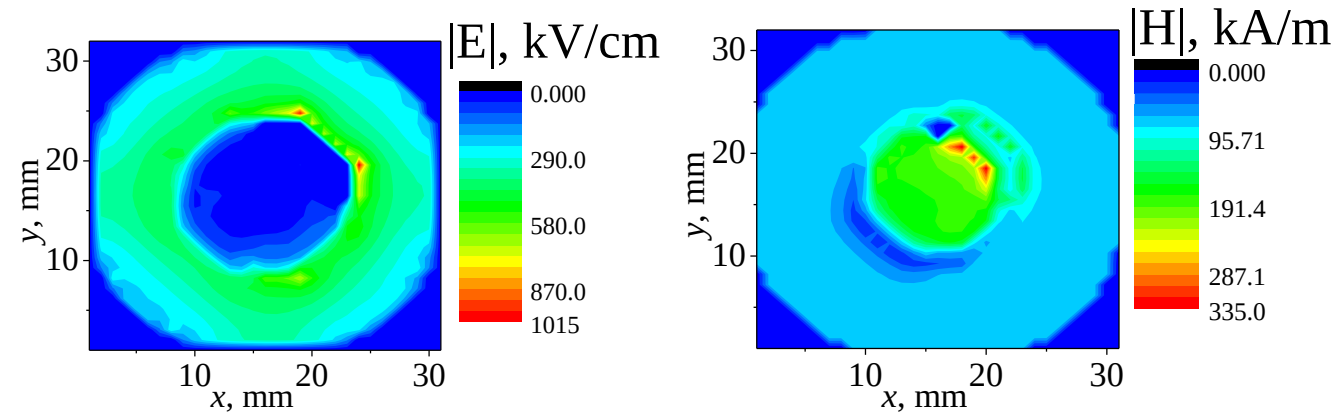

**Fig. 7.** Transverse electric and magnetic field distributions at the center of the spiral conductor design

A significant practical challenge in implementing the proposed spiral GNLTL configuration is the impedance mismatch at the junction between the conventional input feedline and the slow-wave spiral structure. Due to the high characteristic impedance inherent to the helix geometry, direct coupling induces severe pulse reflections, reducing energy transfer efficiency, which is clearly visible in Fig. 6 as a noticeable decrease in the DC component of the output signal compared to the input pulse. To compensate for this effect and achieve optimal impedance matching, we propose systematically increasing the permittivity $\varepsilon_d$ of the background dielectric filling the line. As illustrated in Fig. 8, when the permittivity of the insulating dielectric is decreased to $\varepsilon_d$=1.45 (liquid nitrogen), the impedance mismatch is exacerbated, resulting in even more pronounced pulse reflections. Conversely, increasing the permittivity to $\varepsilon_d$=4 (typical for plastic materials) significantly improves the matching conditions. With a further increase to $\varepsilon_d$=6 (for example, glass), the energy losses caused by the impedance mismatch become nearly negligible.

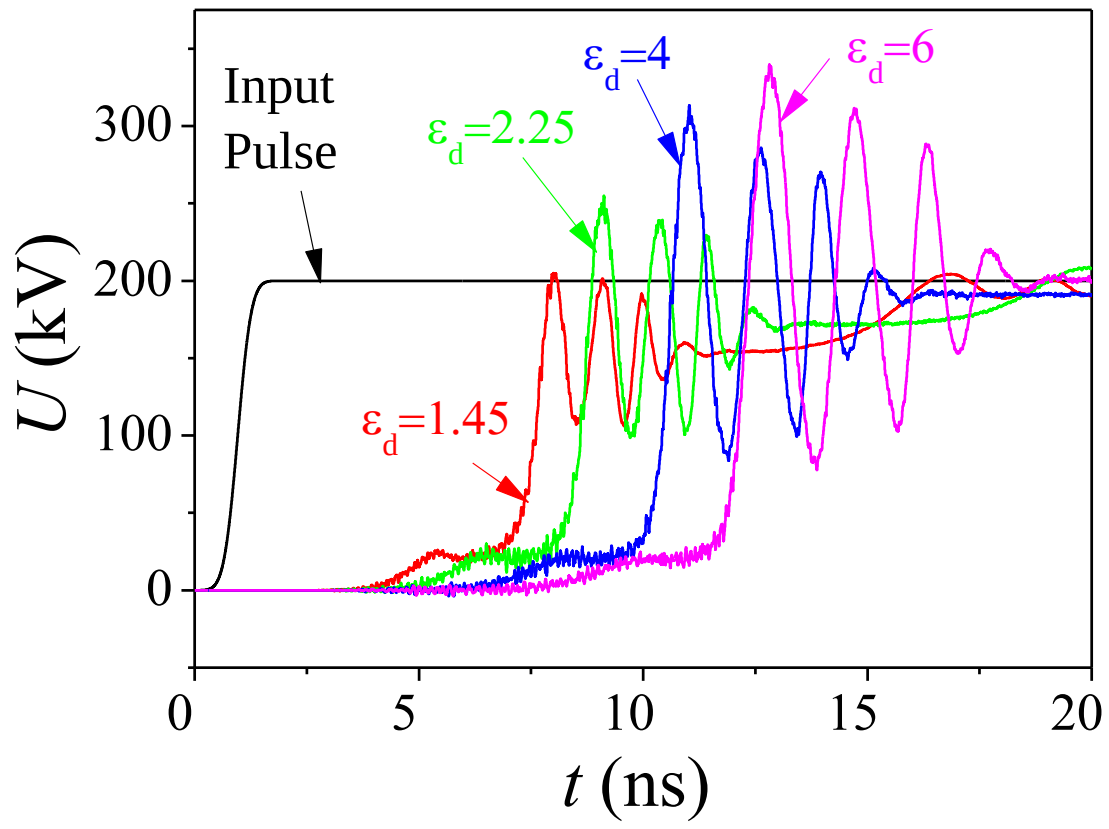

**Fig. 8.** Influence of the insulating dielectric permittivity on the impedance matching observed via output voltage waveforms of the spiral GNLTL configuration calculated for a 200-mm line length.

## VI. Conclusion

In this work, a novel design of a coaxial gyromagnetic nonlinear transmission line featuring a spiral central conductor has been proposed and investigated. To analyze this structurally asymmetric configuration, the specialized 2D axisymmetric FDTD-LLG framework was successfully extended to a fully 3D numerical framework in Cartesian coordinates. The developed 3D computational tool demonstrated stable mesh convergence and satisfactory agreement with the proven 2D baseline model, achieving a modest but acceptable agreement within 5–10% for qualitative verification.

The 3D modeling results confirmed that the proposed spiral configuration successfully replicates the long-lasting, slowly decaying oscillation packet, qualitatively reproducing the previously discovered Cherenkov-like effect observed in GNLTLs with ultra-high-permittivity dielectrics that are impractical for real applications. Furthermore, field profile analysis revealed that the wave slowing effect induces a strong concentration of the magnetic field inside the helix, precisely where the ferrite core is located, thereby enhancing ferrite nonlinearity and improving overall generation efficiency. Finally, the important engineering problem of impedance mismatch at the input of the slow-wave structure was addressed, demonstrating its mitigation through the use of an insulating dielectric with an increased permittivity of 4–6.

## Acknowledgment

The authors acknowledge the use of AI Gemini for proofreading and improving the English phrasing of the manuscript.

## References

[1] J. E. Dolan, “Radio frequency and microwave signal generation by nonlinear magnetic transmission lines,” PCT Int. Patent Appl. WO 2007/141576 A1, Dec. 13, 2007.

[2] J. E. Dolan, “Simulation of shock waves in ferrite-loaded coaxial transmission lines with axial bias,” *J. Phys. D: Appl. Phys.*, vol. 32, no. 15, pp. 1826–1831, Aug. 1999, doi: 10.1088/0022-3727/32/15/310.

[3] J. M. Johnson et al., “Characteristics of a four element gyromagnetic nonlinear transmission line array high power microwave source,” *Rev. Sci. Instrum.*, vol. 87, no. 5, p. 054704, May 2016, doi: 10.1063/1.4947230.

[4] A. J. Fairbanks, A. M. Darr, and A. L. Garner, "A Review of Nonlinear Transmission Line System Design," *IEEE Access*, vol. 8, pp. 148606–148621, 2020, doi: 10.1109/ACCESS.2020.3015715.

[5] W. Hendricks et al., “Experimental study of small-scale coaxial gyromagnetic nonlinear transmission lines,” *IEEE Trans. Plasma Sci.*, vol. 54, no. 3, pp. 1006–1011, Mar. 2026, doi: 10.1109/TPS.2026.3663081.

[6] W. Zhang, M. Lin, H. Li, and X. Qi, "Finite element analysis of pulse sharpening effect of gyromagnetic nonlinear transmission line based on Landau-Lifshitz-Gilbert equation," *Rev. Sci. Instrum.*, vol. 95, no. 6, p. 064705, Jun. 2024, doi: 10.1063/5.0203542.

[7] Y. Cui, J. Meng, Y. Yuan et al., "Principle analysis and preliminary experiment of the gyromagnetic nonlinear transmission lines," *J. Natl. Univ. Defense Technol.*, vol. 46, no. 3, pp. 222–228, Jun. 2024, doi: 10.11887/J.CN.202403022.

[8] Y. Cui, J. Meng, L. Huang and D. Zhu, "100-MW-Level Experiments of a Gyromagnetic Nonlinear Transmission Line System," *IEEE Transactions on Electron Devices*, vol. 69, no. 9, pp. 5248–5255, Sept. 2022, doi: 10.1109/TED.2022.3192215.

[9] A. F. G. Greco et al., "Analysis of the sharpening effect in gyromagnetic nonlinear transmission lines using the unidimensional form of the Landau-Lifshitz-Gilbert equation," *Rev. Sci. Instrum.*, vol. 93, no. 6, p. 065101, Jun. 2022, doi: 10.1063/5.0087452.

[10] Y. Cui et al., "A wideband high-power microwave radiation source based on gyromagnetic nonlinear transmission line and Vlasov antenna," *Rev. Sci. Instrum.*, vol. 93, no. 10, p. 104706, Oct. 2022, doi: 10.1063/5.0102437.

[11] G. J. Deng et al., "Verification of spinwave excitation in coaxial gyromagnetic nonlinear transmission lines," *Rev. Sci. Instrum.*, vol. 94, no. 9, p. 094704, Sep. 2023, doi: 10.1063/5.0155923.

[12] S. Y. Karelin, “FDTD analysis of nonlinear magnetized ferrites: simulation of oscillation forming in coaxial line with ferrite,” *Telecommunications and Radio Engineering*, vol. 76, no. 10. 2017, pp. 873-882. doi: 10.1615/TelecomRadEng.v76.i10.40.

[13] S.Y. Karelin, V.B. Krasovitsky et al., “Quasi-harmonic oscillations in a nonlinear transmission line, resulting from Cherenkov synchronism,” *Probl. At. Sci. Technol.* 2019, vol. 122, no. 4, p. 65-70. DOI: 10.46813/2019-122-065.

[14] S. Karelin and I. Onishchenko, “2D/3D FDTD modeling of coaxial waveguide with magnetized nonlinear ferrite and slow-wave structure,” in *Proc. XXXII Int. Workshop Optical Wave & Waveguide Theory Numerical Modelling (OWTNM26)*, Lausanne, Switzerland, Apr. 2026,

P23. Available: https://owtnm26.epfl.ch/wp-content/uploads/2026/03/OWTNM26_BookOfAbstracts_final.pdf

[15] S. Y. Karelin, V. B. Krasovitsky, I. I. Magda, V. S. Mukhin, and V. G. Sinitsin, "Radio frequency oscillations in gyrotropic nonlinear transmission lines," *Plasma*, vol. 2, no. 2, pp. 258–271, Jun. 2019, doi: 10.3390/plasma2020018.

[16] L. Huang, J. Meng, D. Zhu, and Y. Yuan, "Field-line coupling method for the simulation of gyromagnetic nonlinear transmission line based on the Maxwell-LLG system," *IEEE Transactions on Plasma Science*, vol. 48, no. 11, pp. 3847–3853, Nov. 2020, doi: 10.1109/TPS.2020.3029524.

[17] A. M. Belyantsev and A. B. Kozyrev, "Generation of high-frequency oscillations by electromagnetic shock wave on transmission lines on the basis of multilayer heterostructures," *Tech. Phys.*, vol. 42, no. 6, pp. 615–620, Jun. 1997, doi: 10.1023/A:1022615411450.

[18] P. D. Coleman, J. J. Borchardt, J. A. Alexander, J. T. Williams, and T. F. Peters, "Characterization of a synchronous wave nonlinear transmission line," in *Proc. 2011 IEEE Pulsed Power Conf. (PPC)*, 2011, pp. 173–177. Available: https://www.osti.gov/servlets/purl/1107564.

[19] N. Seddon, C. R. Spikings, and J. E. Dolan, "RF pulse formation in nonlinear transmission lines," in *2007 16th IEEE Int. Pulsed Power Conf.*, vol. 1, Jun. 2007, pp. 678-681. doi: 10.1109/PPPS.2007.4651931.

[20] A. Taflove and S. C. Hagness, *Computational Electrodynamics: The Finite-Difference Time-Domain Method*, 3rd ed. Norwood, MA, USA: Artech House, 2005.

[21] D. M. French and B. W. Hoff, "Spatially dispersive ferrite nonlinear transmission line with axial bias," *IEEE Transactions on Plasma Science*, vol. 42, no. 10, pp. 2822–2825, Oct. 2014, doi: 10.1109/TPS.2014.2348492.

[22] S. Y. Karelin, V. B. Krasovitsky, and I. I. Magda, "Cherenkov effect and narrow-line VHF oscillations in a ferrite-filled coaxial line," in *Proc. 2022 IEEE 2nd Ukrainian Microwave Week (UkrMW)*, Kharkiv, Ukraine, 2022, pp. 92–95, doi: 10.1109/UkrMW58013.2022.10037087.

[23]. A. B. Batrakov, S. Y. Karelin, O. M. Lebedenko, V. S. Mukhin, I. N. Onishchenko, O. L. Rak, and M. V. Volovenko, "Electromagnetic waves excitation by a unipolar high-voltage pulse in a ferrite coaxial line with a periodic metal structure," *Problems of Atomic Science and Technology*, vol. 155, no. 1, pp. 65–68, 2025, doi: 10.46813/2025-155-065.